\documentclass[twocolumn,letterpaper]{IEEEAerospaceCLS}  

\usepackage[]{graphicx,float,latexsym,amsfonts,bm,amstext,times}
\usepackage{tikz}
\usepackage{copyrightbox}
\usepackage[utf8]{inputenc}
\usepackage{pgfplots} 
\usepackage[export]{adjustbox}
\usepackage{pgfgantt}
\usepackage{pdflscape}
\pgfplotsset{compat=newest} 
\pgfplotsset{plot coordinates/math parser=false}
\usepackage{amssymb,subcaption}
\usepackage{siunitx}
\usepackage[inline]{enumitem}
\usepackage[short]{optidef}

\usepackage[]{graphicx}    

\newcommand{\todo}[1]{\textcolor{red}{TODO: #1}}

\begin{document}
\title{Optimal Attitude Control of Large Flexible Space Structures with Distributed Momentum Actuators}

\author{%
Pedro Rocha Cachim\\ 
Carnegie Mellon University\\
pcachim@andrew.cmu.edu
\and 
Will Kraus\\
Carnegie Mellon University\\
wkraus@andrew.cmu.edu
\and 
Zachary Manchester\\
Carnegie Mellon University\\
zmanches@andrew.cmu.edu
\and 
Pedro Louren\c co\\
GMV\\
palourenco@gmv.com
\and 
Rodrigo Ventura\\
Instituto Superior T\'ecnico\\
rodrigo.ventura@isr.tecnico.ulisboa.pt
\thanks{\footnotesize 979-8-3503-5597-0/25/$\$31.00$ \copyright2025 IEEE}              
}

\maketitle

\thispagestyle{plain}
\pagestyle{plain}

\maketitle

\thispagestyle{plain}
\pagestyle{plain}
\begin{abstract}
Recent spacecraft mission concepts propose larger payloads that have lighter, less rigid structures. For large lightweight structures, the natural frequencies of their vibration modes may fall within the attitude controller bandwidth, threatening the stability and settling time of the controller and compromising performance. This work tackles this issue by proposing an attitude control design paradigm of distributing momentum actuators throughout the structure to have more control authority over vibration modes. The issue of jitter disturbances introduced by these actuators is addressed by expanding the bandwidth of the attitude controller to suppress excess vibrations.
Numerical simulation results show that, at the expense of more control action, a distributed configuration can achieve lower settling times and reduce structural deformation compared to a more standard centralized configuration.
\end{abstract}

\tableofcontents

\section{Introduction}

\begin{figure}[]
        \centering
        \begin{subfigure}[b]{0.23\textwidth}
            \centering
            \includegraphics[width=\textwidth]{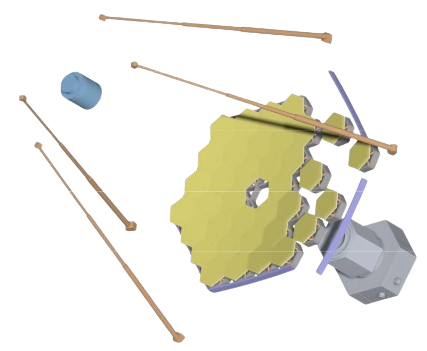}
            \caption[AAST~\protect\cite{basu2003proposed}]%
            {{\small AAST~\protect\cite{basu2003proposed}}}    
            \label{fig:aast}
        \end{subfigure}
        \hfill
        \begin{subfigure}[b]{0.23\textwidth}  
            \centering 
            \includegraphics[width=\textwidth]{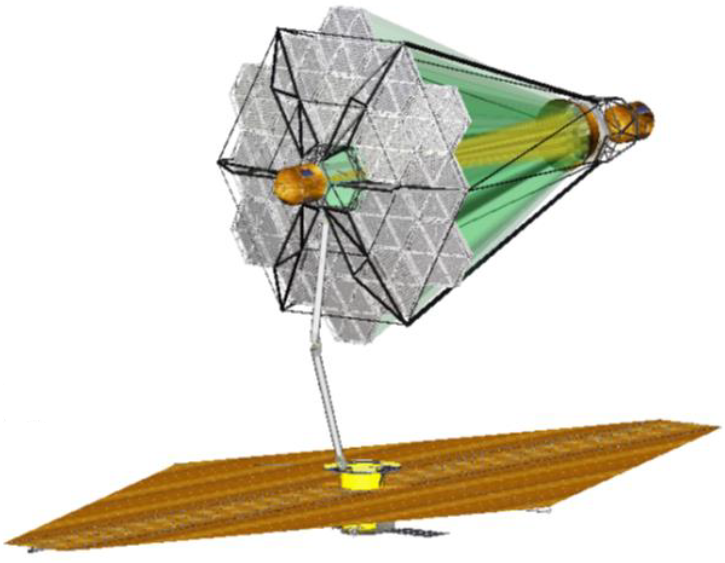}
            \caption[EST~\protect\cite{polidan2015evolvable}]%
            {{\small EST~\protect\cite{polidan2015evolvable}}}    
            \label{fig:est}
        \end{subfigure}
        \vskip\baselineskip
        \begin{subfigure}[b]{0.20\textwidth}   
            \centering 
            \includegraphics[width=\textwidth]{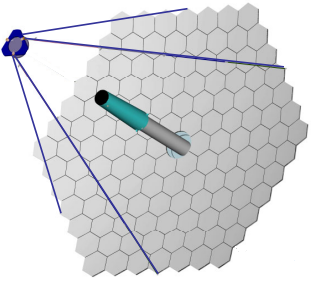}
            \caption[NNGST~\protect\cite{muller2002assembly}]%
            {{\small NNGST~\protect\cite{muller2002assembly}}}    
            \label{fig:nngst}
        \end{subfigure}
        \hfill
        \begin{subfigure}[b]{0.20\textwidth}   
            \centering 
            \includegraphics[width=\textwidth]{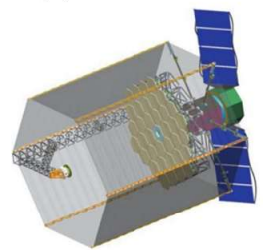}
            \caption[OAST~\protect\cite{boqian2020conceptual}]%
            {{\small OAST~\protect\cite{boqian2020conceptual}}}    
            \label{fig:ramst}
        \end{subfigure}
        \caption[Very Large Space Telescope Concepts \protect\cite{kuang2024design}.]
        {\small Very Large Space Telescope Concepts illustrating the growing demands for larger-aperture space telescopes from the space science community \cite{kuang2024design}.} 
        \label{fig:telconc}
\end{figure}

Present and future generations of earth observation, space science, and telecommunication spacecraft are pushing the boundaries of pointing performance, payload mass, and cost. Finer pointing enables increased detector resolution, sensitivity and longer integration time \cite{sanfedino2022advances}, and higher gains can be achieved with a wider or longer antenna or a larger telescope aperture. Several mission concepts, such as the ones shown in Fig. \ref{fig:telconc}, have been proposed \cite{kuang2024design}.
Mission cost is the determining factor for the feasibility of these proposals, which quickly scales with the mass of the system. Mass can be reduced at the expense of rigidity, which can push vibration modes to lower frequencies, potentially falling within the bandwidth of the attitude control system (ACS) and complicating pointing \cite{liu2024review}.

Traditional ACS architectures have momentum actuators such as Reaction Wheels (RW) and Control Moment Gyros (CMG) centralized on the bus, reducing controllability over the spacecraft (SC) rotational degrees of freedom (DoF) as it loses rigidity.
Previous studies have analyzed the advantages of distributing attitude control momentum actuators to increase the control authority over the non-rigid dynamics, concluding that these changes improve the damping of the vibrations in the system and lower their settling time \cite{bifa2024multi,hu2020flexible}. 



To the authors' knowledge, no previous study has addressed the main design drivers that have led to the preference for few centralized large actuators in previous missions. For high-accuracy space-science missions, a primarly driver is the jitter perturbations induced by these actuators, which scale quadratically in magnitude and linearly in frequency with wheel speed. The common approach in the aerospace industry to reduce this behavior is to have large wheels operating at low speeds around narrow operating ranges \cite{hasha2016high}.

\begin{figure}[]
\centering
\includegraphics[width=1.0\linewidth]{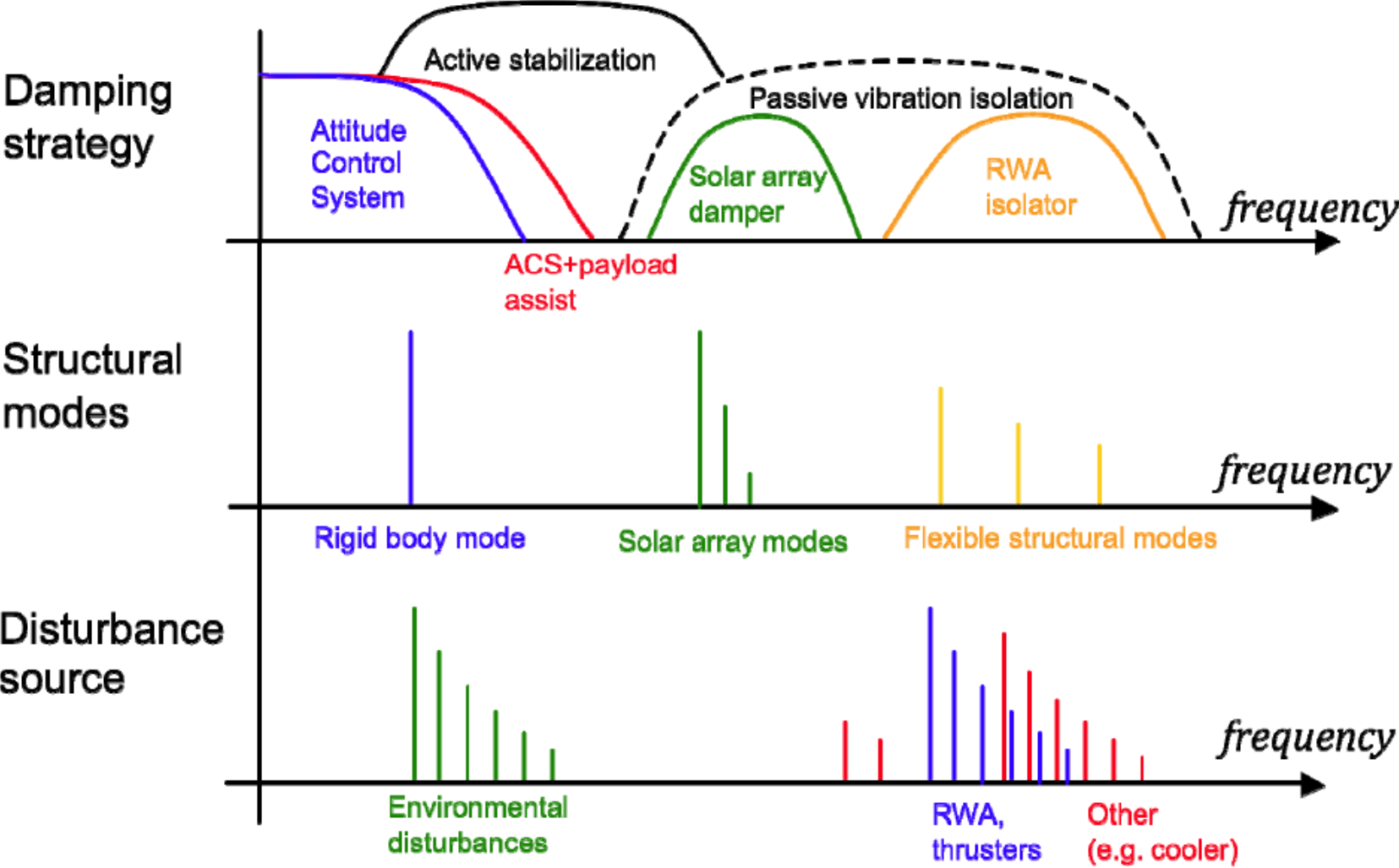}
\caption{Notional frequency spectrum of spacecraft disturbances (bottom), rigid and flexible body modes (middle) and typical vibration suppression strategies (top) \protect\cite{blackmore2011instrument}.}
\label{fig:bdwths}
\end{figure}

This approach becomes more challenging to implement as structures become less rigid and the structural response expands to lower frequencies \cite{preda2018robust}. The consideration of these behaviors in the attitude control design, or even the inclusion of a dedicated vibration suppression system, help minimize these undesirable dynamics. A notional representation of the frequency spectrum of these systems is shown in Fig. \ref{fig:bdwths}. Vibration suppression systems are designed to suppress vibrations on a low-to-medium frequency range, typically from a few \SI{}{\Hz} up to a few hundred \SI{}{\Hz} above the attitude control system bandwidth, and this covers the frequency of the main RW harmonic disturbance \cite{preda2017robust}. However, in some scenarios, these mechanisms have been found to be inefficient in suppressing large-amplitude, low-frequency behavior \cite{tang_-orbit_2020,hu2020flexible}, which is more pronounced with larger and less rigid structures. In addition, they are commonly deployed in parallel with the ACS, acting as competitors and canceling out the disturbing effects from the other \cite{balas1979direct,fanson1990positive,wie1988pole}. The concept of merging both the vibration suppression systems and the ACS into a single system has been explored in recent literature, with results showing similar performances using less control cost compared to a separated system but with a tradeoff of more complex controller design \cite{angeletti2021end,sabatini2020synergetic}. 



In this work, the performance of a distributed RW actuator approach for attitude control and vibration suppression is compared with the standard centralized approach. 
The main contributions of this research are the merging of the previous research on synergetic ACS/vibration suppression systems and the distributed momentum actuator approach and the consideration of RW disturbances in the system design.
The ACS system's operational bandwidth is extended to cover the typical vibration suppression frequencies and reject the disturbances, merging the work from previous studies on synergetic ACS/vibration suppression systems and distributed ACS momentum actuator control for very large structures. 
The controllers for the distributed and centralized systems are designed to optimize a similar quadratic cost function.
Numerical simulation results show that the distributed version yields a more agile attitude controller with less structural deformation at the cost of more control action.


The paper proceeds as follows: Section \ref{sec:sysmodel} describes the dynamic model of the flexible space structure used as a case study as well as the sensors and actuators of the system.
Section \ref{sec:ctrlsys} describes the design and architecture of the ACS for the case study.
In Section \ref{sec:simres}, the simulation results are shown and discussed. Finally, in Section \ref{sec:conc}, conclusions are drawn on the performance of the distributed approach versus the centralized approach as well as the use of the ACS for vibration suppression.  

\section{System Modeling}\label{sec:sysmodel}
\begin{figure*}
\centering
\begin{subfigure}{.48\textwidth}
  \centering
  \includegraphics[width=\linewidth]{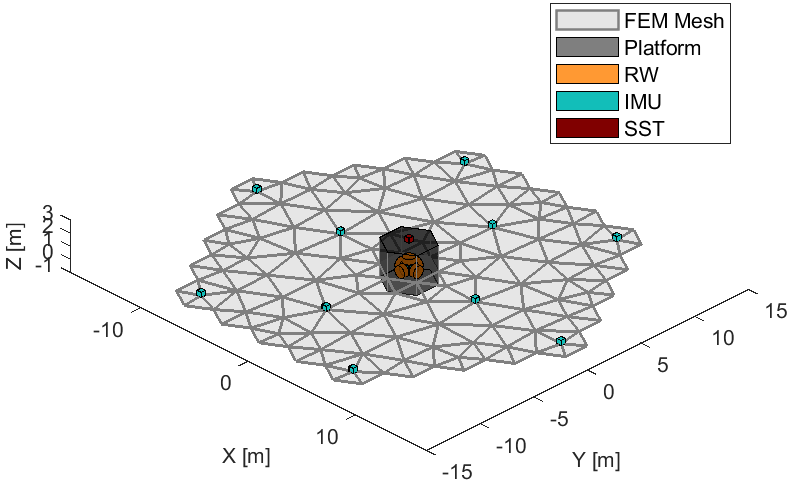}
  \caption{Centralized configuration. Six large momentum actuators are placed on the spacecraft bus in a cubic hexahedral arrangement, with vibration sensors distributed throughout the structure.}
  \label{fig:CentSC}
\end{subfigure}%
\hfill
\begin{subfigure}{.48\textwidth}
  \centering
  \includegraphics[width=\linewidth]{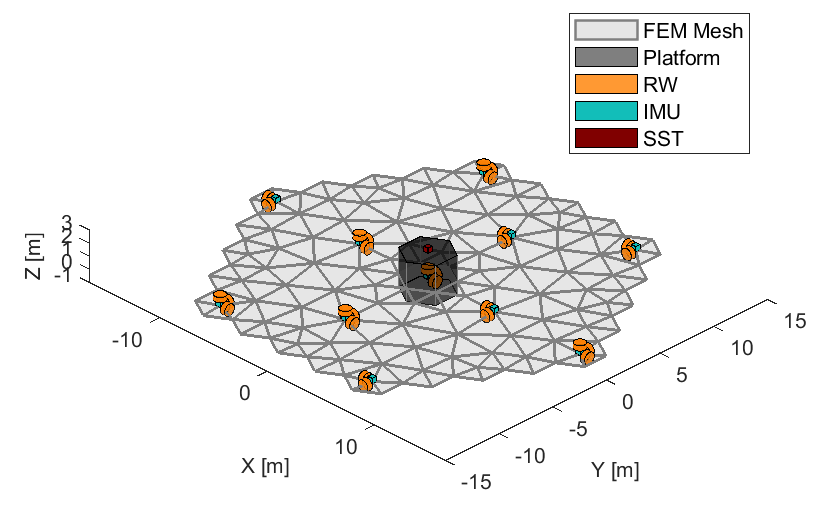}
  \caption{Distributed configuration. 11 sets of three momentum actuators are placed throughout the entire structure, including the bus, with collocated vibration sensors.}
  \label{fig:DistSC}
\end{subfigure}
\caption{Centralized and distributed actuator attitude control system architectures. Both scenarios have a similar total angular momentum capacity and maximum torque in all axis, only differing in the location of the actuators.}
\label{fig:SCModel}
\end{figure*}

The flexible structure model for this case study shown in Fig. \ref{fig:CentSC} and \ref{fig:DistSC} is based on the very large telescope concepts shown in Fig.~\ref{fig:telconc}. This model is composed of a central rigid spacecraft platform, a large flexible structure attached to the platform, reaction wheel actuators with Inertial Measurement Unit (IMU) sensors and a star tracker (SST) for attitude determination. 
The parameters of the system are described in Table \ref{tab:simparams}.
The Finite Element Analysis (FEA) mass, stiffness and modal participation matrices $M_{\mathcal{N}}$, $K_{\mathcal{N}}$ and $L_{\mathcal{N}}$ of the system were obtained from simplified discrete Kirchhoff plate elements \cite{tian2021triangular}. Kirchhoff plate theory considers the approximation of bending only without shear forces. This assumption is generally valid for plates under small displacements that have a side-to-thickness ratio greater than 30 \cite{reddy2006theory}. 
The nodal degrees-of-freedom (DoFs) related to torsion, compression and extension are neglected.

\begin{figure*} 
\centering
\includegraphics[width=1.0\linewidth]{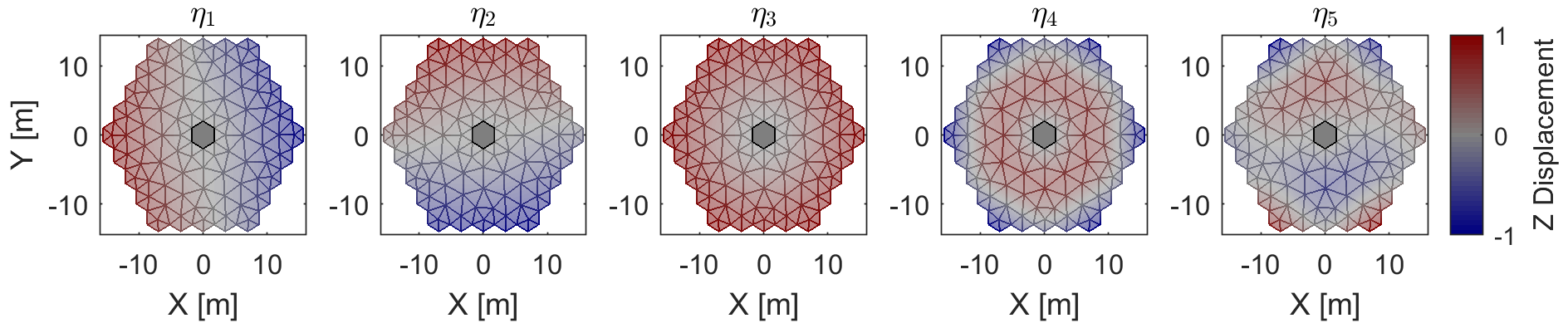}
\caption{Five lowest vibration frequency mode shapes of the flexible structure. The first two modes are strongly coupled to the rigid body rotation of the structure around the X and Y axis.}
\label{fig:scmodes}
\end{figure*}

The flexible structure is cantilevered to the rigid platform through the inner 6 nodes, such that the linear and rotational DoFs of those nodes are set to zero. 
Once clamped, the FEM plate has 396 nodal DoFs. To facilitate numerical analysis, model order reduction is performed via modal truncation. This is performed through the mass-normalized modal matrix $\Phi=\begin{bmatrix}\phi_1 & \phi_2 & \ldots & \phi_n \end{bmatrix}$ of the system, obtained from the eigenvectors of the stiffness matrix divided by the mass matrix:
\begin{equation}
    \begin{cases}
        (K_{{\mathcal{N}}}-\omega_i^2M_{{\mathcal{N}}})\phi_i &= 0 \\ 
        M=\Phi^TM_{{\mathcal{N}}}\Phi &= I\\
        K=\Phi^TK_{{\mathcal{N}}}\Phi &= \text{diag}(\omega_1^2, \ldots, \omega_n^2)\\ 
    \end{cases}
    \label{eq:nod2mode}
\end{equation}
where $\bm{\phi_i}$ is a linear combination of nodal displacements that form a mode shape, associated with the modal frequency $\omega_i$. The system can then be reduced by truncating out higher frequency modes, keeping only a selection of the lowest frequency ones. The model is reduced to 25 frequency modes below a frequency of $80\SI{}{\Hz}$ with the strongest coupling to the rigid body modes. The five lowest of these are shown in Fig. \ref{fig:scmodes}. The truncated mass-normalized modal matrix $\Phi_t$ approximates the relation between the modal and mesh nodal coordinates $\eta=\Phi^{-1}q_{{\mathcal{N}}}=\Phi^{T}M_{{\mathcal{N}}}q_{{\mathcal{N}}}$ such that $\eta\approx\Phi_t^TM_{{\mathcal{N}}}q_{{\mathcal{N}}}$.

The equations of motion of the system are described as:
\begin{equation}
    \begin{cases}
        \dot{p} = \dfrac{1+p^Tp}{4}\left(I_3 + 2 \dfrac{[p\times]^2+[p\times]}{1+p^Tp}\right)\omega \\
        m \dot{v} + L_{v} \ddot{\eta} + \omega \times m v= 0 \\
        J \dot{\omega} + L_{\omega} \ddot{\eta} + G_{\omega}(\omega,\dot{\eta},\rho_{rw})= B_{\omega} \tau \\
        L_{v}^T \dot{v} + L_{\omega}^T \dot{\omega} + \ddot{\eta} + F \dot{\eta} + K \eta + G_{\eta}(\omega,\dot{\eta},\rho_{rw}) = B_{\eta} \tau \\
        \dot{\rho}_{rw} = \tau
    \end{cases},
    \label{eq:sysdyn}
\end{equation}
where $p$ are the Modified Rodrigues Parameters (MRP), $[p\times]$ represents the skew-symmetric matrix encoding the cross-product, $J$ is the moment of inertia of the SC, $\omega$ is its inertial to body frame angular velocity expressed in the body frame, $v$ is the body-frame linear velocity, $m$ is the mass of the system, $\rho_{rw}$ is the angular momentum of the RWs, $B_{\omega}$ and $B_{\eta}$ are the input jacobians, $G_{\omega}$ and $G_{\eta}$ are the nonlinear gyricity vectors and $\tau$ is the torque applied on the RW. The structural damping of the system is modeled with the standard Rayleigh damping approach, with $F=\beta_M M + \beta_K K$.

The input jacobians $B_{\omega}$ and $B_{\eta}$ relate the reaction torque of the RWs on the structure rigid rotational and flexible states, described by the direction of the RW rotation axis in the body frame and the deformation slope where the RW is placed as a function of the modal coordinates.
The gyricity vectors of the system $G_{\omega}$ and $G_{\eta}$ are defined as in previous literature \cite{ploen2004rigid}, accounting for the gyroscopic torque of the RW with the angular velocity of the SC and the local deformation of the structure.

The measurements $y$ of the system consist of the IMU measurements, which uses the accelerometer (AMU) $y_{amu}$, gyrometer $y_{gyr}$, RW encoders $y_{rw}$ and SST, is defined as:

%
\begin{align}
    y_{sst} &= p; \quad y_{gyr} = \omega + C_{gyr}\dot{\eta}\\ 
    y_{amu} &= \dot{v} + \dot{\omega}\times r_{amu}+ 2 \omega \times C_{amu}\dot{\eta} \\ &+ \omega \times \omega \times r_{amu} + C_{amu}\ddot{\eta} \\
    y_{rw}  &= \omega_{rw} - B_{\omega}^T\omega - C_{rw}\dot{\eta}
\end{align}
where $C_{gyr}$,  $C_{amu}$ and $C_{rw}$ relate the modal coordinates to the local slope deformation for the gyrometers and reaction wheels and to the local displacement for the AMUs. A second-order low-pass Butterworth filter is applied on the setup with cutoff frequency at \SI{80}{\Hz} for anti-aliasing. 

\subsection{Disturbance Model}\label{sec:distmodel}

The wheel imbalance disturbances are modeled as exogenous forces and torques applied on the structure. These forces and torques are described in the actuator frame (with the z axis aligned with the axis of rotation) as follows \cite{kim2014micro}:
\begin{align}
    f_{x}(t) &= \sum_{i=1}^{n_f}C_i^f\omega_{rw}^2\sin(h_i^f\omega_{rw} t + \phi_i^f)\\
    f_{y}(t) &= \sum_{i=1}^{n_f}C_i^f\omega_{rw}^2\cos(h_i^f\omega_{rw} t + \phi_i^f)\\
    \tau_{x}(t) &= -\sum_{i=1}^{n_{\tau}}C_i^{\tau}\omega_{rw}^2\cos(h_i^{\tau}\omega_{rw} t + \phi_i^{\tau})\\
    \tau_{y}(t) &= \sum_{i=1}^{n_{\tau}}C_i^{\tau}\omega_{rw}^2\sin(h_i^{\tau}\omega_{rw} t + \phi_i^{\tau})
\end{align}
where $C_i^f$ and $C_i^{\tau}$ are the force and torque coefficients for the harmonic $i$, $h_i^f$ and $h_i^{\tau}$ are the force and torque harmonic orders and $\phi_i^f$ and $\phi_i^{\tau}$ are the phases of the force and torque perturbations. Also, the simplifying assumptions from \cite{preda2018robust} are also considered: $\phi_i^f=\phi_i^{\tau}$, only lateral forces and torques on each wheel orthogonal to the axis of rotation are considered and all harmonics beyond the first order are modeled as a broadband white noise.
The scaling of these disturbances with respect to the moment of inertia between the large RW in the centralized configuration and the small RWs used in the distributed configuration is considered linear.

\subsection{Linearized Dynamics}\label{sec:lindyn}

The described model can be linearized around a stationary point of operation. In a small enough neighborhood of this stationary point within the state space, the dynamics may be approximated to this linearized model. Assuming always a small enough SC angular velocity, the system is linearized around a nominal RW angular velocity/momentum. The only non-linear terms in the dynamics in Eq. \ref{eq:sysdyn} are expressed by the gyric vector, which can be linearized as stated in \cite{ploen2004rigid}:
\begin{equation}
    \dfrac{dG_{\omega,\eta}}{d\omega}\Bigr|_{\rho_0} = \begin{bmatrix}
    \left[\rho_0\times\right] \\ B_{\eta}\left[\rho_0\times\right]
\end{bmatrix}, \dfrac{dG_{\omega,\eta}}{d\dot{\eta}}\Bigr|_{\rho_0} = \begin{bmatrix}
    \left[\rho_0\times\right]B_{\eta}^T \\ B_{\eta}\left[\rho_0\times\right]B_{\eta}^T
\end{bmatrix}
\end{equation}
and the kinematics of the MRP, which are linearized around a null angular velocity such that $\delta \dot{p}\approx \delta \omega/4$.

In the measurement equations, all second order terms for the accelerometer are cancelled in the linear approximation, such that
\begin{equation}
    y_{amu} \approx \delta\dot{v} + \delta\dot{\omega}\times r_{amu} + C_{amu}\delta \ddot{\eta}.
\end{equation}
The disturbance model is linearized such that the magnitude is constant and based on the reference angular velocity of the wheels for the given scenario.

\section{System Design}\label{sec:ctrlsys}

\begin{figure}
\centering
\includegraphics[width=1.0\linewidth]{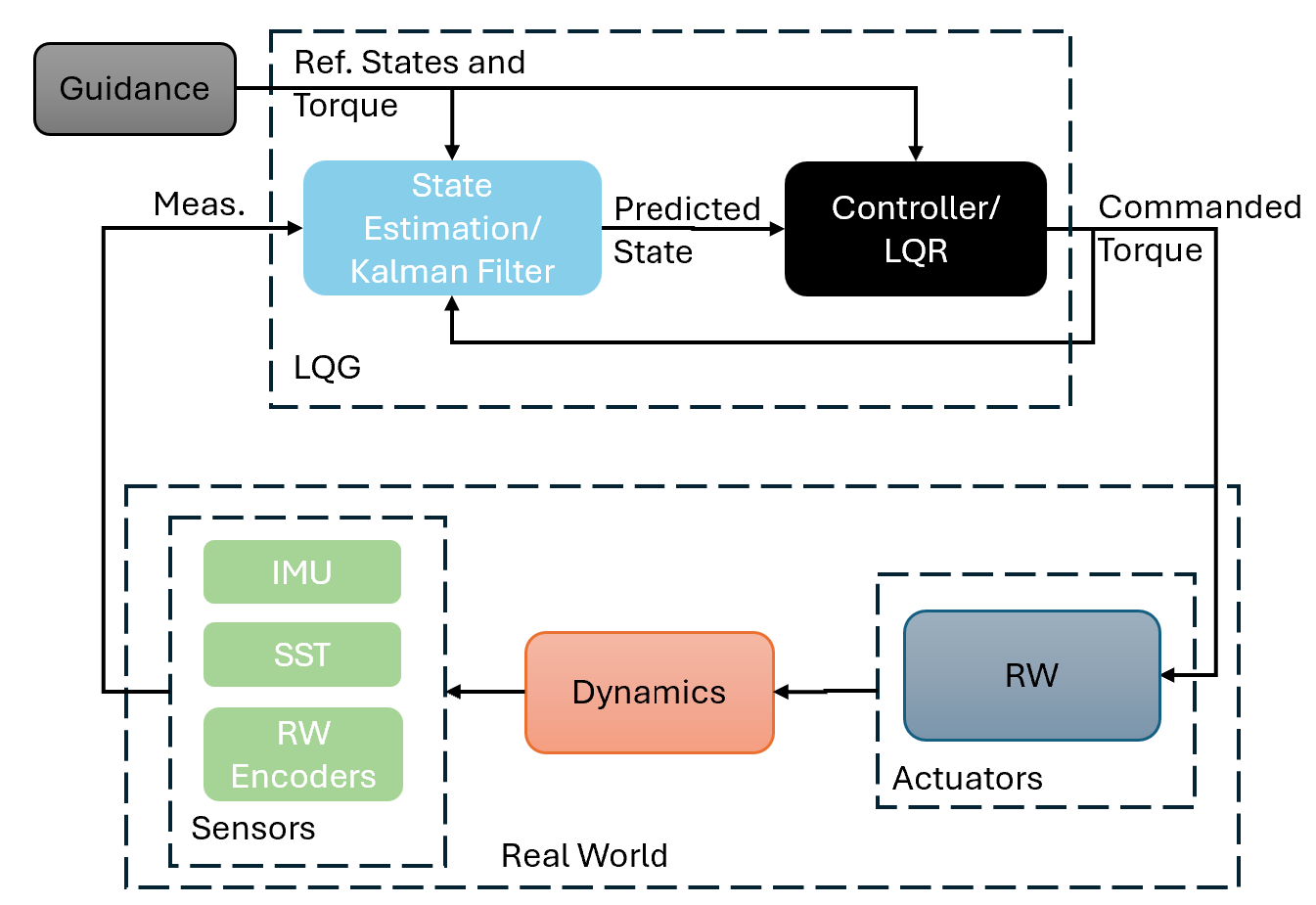}
\caption{High-level system architecture for the flexible structure controller.}
\label{fig:method}
\end{figure}

To achieve stable pointing with accurate reference tracking, the integrated approach shown in Fig. \ref{fig:method} is proposed. Two algorithms are designed as part of the ACS for the case study: an attitude guidance algorithm and a low-level, high-frequency joint state controller with state estimator for disturbance rejection.




\subsection{Actuator/Sensor Placement}

The distribution of actuators and sensors along the structure is a key factor in the design of vibration control systems. A well-known design preference is the collocation of actuators and sensors with direct feedback velocity, offering robustness to the closed loop stability \cite{lee2011comparison,verma2020perfect}.

The distributed actuation configuration considers 11 sets of three collocated sensors/actuators aligned with the SC body frame XYZ axis, as shown in Fig. \ref{fig:DistSC}. The centralized actuation configuration considers two of those sets in the satellite bus to cancel out the system's total angular momentum when the wheels are at their nominal speed, which is illustrated in Fig. \ref{fig:CentSC}.
The distributed configuration actuators are downsized with respect to the actuator used in the centralized configuration to each have $2/11$ths of the maximum torque and inertia (the dimensions are provided in Table \ref{tab:simparams} in Appendix \ref{app:sysparam}).
One set of sensors is placed on the platform, together with a SST to offer full observability of the rigid body rotation states. The remaining 10 sets of sensor-actuator pairings are placed evenly around the flexible structure. Due to redundancy, the angular rate sensor around the Z axis and the accelerometers on the X and Y in-plane are removed. Due to the model assumptions, the angular rate sensor around the Z axis and the accelerometers on the X and Y in-plane do not capture the local vibrations of the plate and are therefore not included. The optimization of the distribution of the actuator-sensor pairings is left for a future study.

\subsection{RW Wheel Speed Constraints}
One key factor in system design affected by the actuator-sensor pairing placement is the selection of the RW operational speed range. This step is critical for minimizing the effect of the RW disturbances on the controller performance.
The natural frequencies that an actuator is able to excite depends on its location. If the mode shape has a pronounced slope around the main rotation axis of the RW, then the natural frequency will be strongly coupled to the RW actuation. The same applies to the harmonic static and dynamic imbalances of the wheels and the direction/location of the forces/torques they produce along the flexible structure. Therefore, selecting a range of speeds such that the magnitude of the structural displacement is minimal is desirable.

A trade-off must be considered when selecting the allowed bandwidth. The choice of RW speed constraints also defines their angular momentum storage capacity and the achievable slew rate: the tighter the constraints, the slower the SC will be able to slew. Additionally, as the RW speed nears the imposed limits to avoid structural interference, desaturation maneuvers must be performed to reset the wheel speed back to the nominal configuration. These maneuvers will be needed more frequently if the allowed bandwidth of the RW actuators is tightened, posing a potential risk to operational performance. The allowed speed range for the RW is 50 rpm around the nominal speed. For the sake of simplicity, the total angular momentum storage ability is considered similar in both the centralized and distributed scenarios. This ability could be optimized with the strategic placement of the actuators. In addition, to avoid stiction at zero-crossings \cite{rigger2010stiction}, a minimum angular velocity of $150$ rpm is defined for each wheel. Due to the tight RW speed constraints, the maximum slew speed for this case study mission is $\approx \sim 0.25$ degree per minute for rotations around the $x$ and $y$ body axis. Typical telescope missions hold slew rates in the order of a few degrees per minute, tending towards lower values for larger missions \cite{ponche2023guidance}. A mission of comparable moment of inertia to this case study, ATHENA \cite{ATHENA}, is planned to hold a nominal slew rate of $\sim 1 $ degree per minute. ATHENA, however, has double the mass of this system and is both smaller and more rigid than the SC in this case study.

\begin{figure}
\centering
\includegraphics[width=1.0\linewidth]{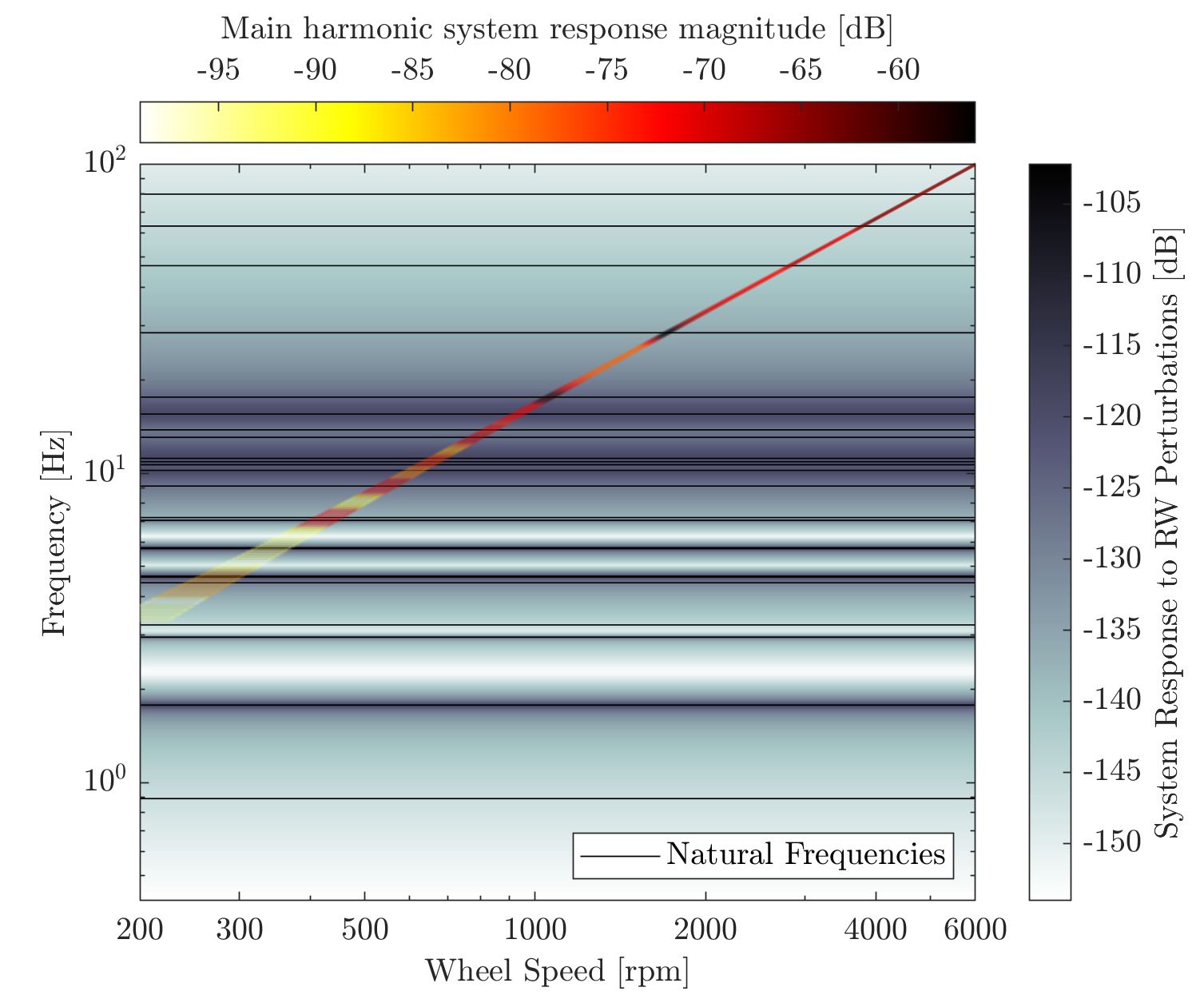}
\caption{System response spectrum as a function of the wheel speed and magnitude of the main harmonic disturbance. The frequency response of the system is not significantly affected by the change in reaction wheel speed.}\label{fig:rwdist}
\end{figure}

The nominal rotation velocity of each wheel was defined separately. In previous studies, models of the passive damping on the wheels were considered that accounted for the change in the structural response of the system due to the change in RW angular velocity - the so-called whirl modes \cite{preda2018robust}. In this study, no passive dampers were considered to narrow the scope of the study and to reduce complexity, although the structural response of the system is still affected by the change in RW speed due to the gyricity of the wheels not being neglected. Fig. \ref{fig:rwdist} shows the infinity-norm of the transfer function of the linearized system $G(s)_{w_{rw}\rightarrow \dot{\eta}}$, from the disturbances of all RWs to the modal velocities of the system as a function of all of the RW speed, if they were all spinning around the same nominal velocity. The blue curve shows that, for a safe band around the nominal wheel speed, the peak magnitude of  $\|G(s)_{w_{rw}\rightarrow \dot{\eta}}\|_{\infty}$ is within that band. From this information, the effect of the RW speed on the structural response is found to be minimal, which allows for the optimization of the speed of each RW separately without factoring in the interaction between the different wheel speeds. 

\begin{figure*}
\centering
\begin{subfigure}{.48\textwidth}
  \centering
  \includegraphics[width=\linewidth]{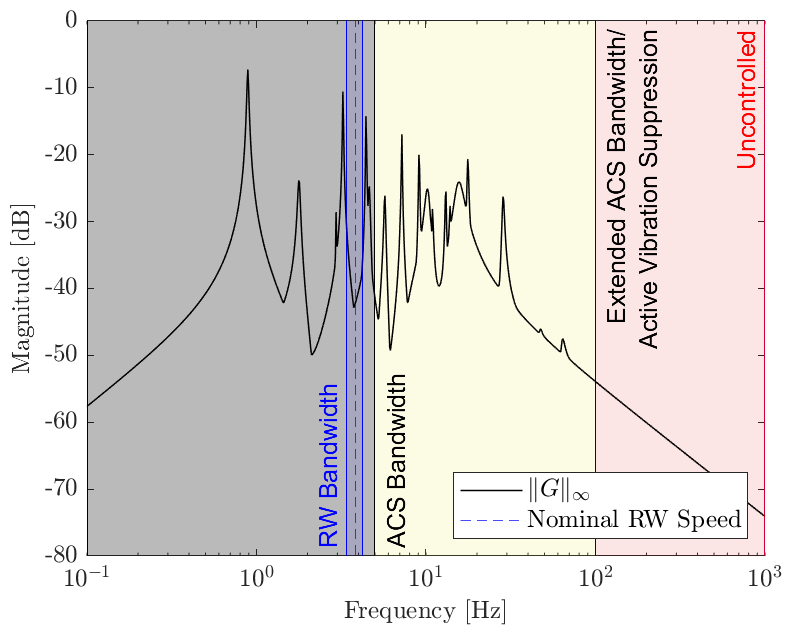}
  \caption{Frequency response of the modal velocities $\dot{\eta}$ to the perturbations of the bus +X facing RW $\|G_{w_{+x}\rightarrow \dot{\eta}}\|_{\infty}$ and optimized RW speed box constraint.}
  \label{fig:CentPSD}
\end{subfigure}%
\hfill
\begin{subfigure}{.48\textwidth}
  \centering
  \includegraphics[width=\linewidth]{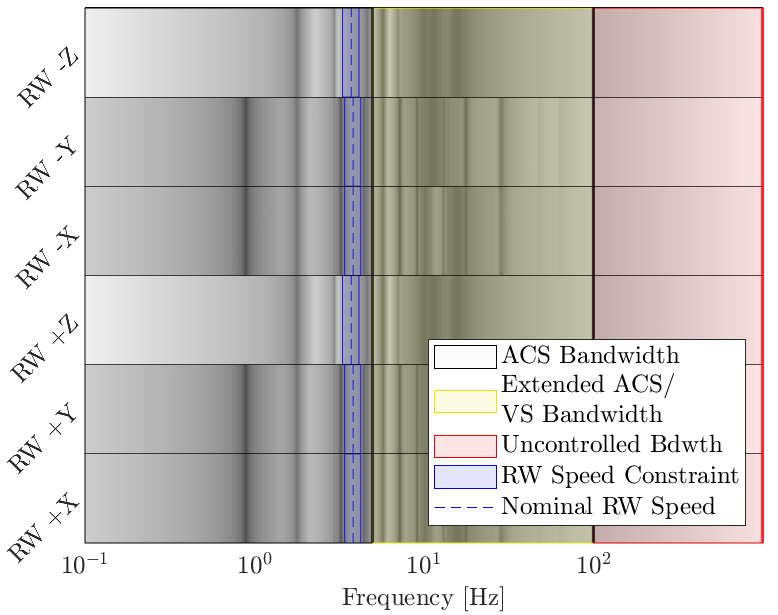}
  \caption{Frequency response of the modal velocities $\dot{\eta}$ to the perturbations of the bus RWs $\|G_{w_{\cdot}\rightarrow \dot{\eta}}\|_{\infty}$ and optimized RW speed box constraint for each.}
  \label{fig:DistPSD}
\end{subfigure}
\caption{Bandwidth of the controller, largest singular value of the MIMO system at the nominal RW speed and Reaction Wheel allowed range.}\label{fig:CentDistPSD}
\end{figure*}

In Fig. \ref{fig:CentDistPSD}, the results of the optimization of the wheel speed range for the centralized scenario are shown as an example. In Fig. \ref{fig:DistPSD}, the result is shown separately for each RW. In Fig. \ref{fig:CentPSD}, the optimized bandwidth is shown only for the wheel in the bus facing the +x direction. This optimization accounts for the disturbance model and the quadratic growth of the forces and torques with the wheel angular velocity, which drives the selection to lower speeds to minimize the energy of the rotations and is consistent with previous studies \cite{le2017micro}. 
\raggedbottom

\subsection{Attitude Guidance} 

In this work, two comparison scenarios requiring guidance/planning reference state profiles are considered: a long slew procedure and wheel off-loading. The reference state and torque profiles are the solution to the following optimization problem:
\begin{mini}
  {\delta x,u}{\dfrac{1}{2}\sum^{N}_{k=1} \|\delta x_k - \delta x_{ref}\|_{Q}}{}{}
  \addConstraint{\delta x_{k}}{=A\delta x_{k-1}+Bu_{k-1},}{k=1,\ldots,N}
  \addConstraint{\|H_i^T\delta \tilde{\omega}_{rw}\|}{\leq \delta \omega_{max},\quad}{k=1,\ldots,N}
  \addConstraint{\|H_i^T u_k\|}{\leq \tau_{max},\quad}{k=0,\ldots,N-1}
\end{mini}
This formulation considers a simplified dynamic model without translational states, flexible modes or the RW null space. The reduced guidance state vector $\delta x_k$ considers only attitude (MRP) $\delta p$, angular velocity $\delta \omega$, the inertial subspace of the RWs of the system and the corresponding torque on the three axis this subspace inputs to the spacecraft and RWs. The input torque subspace that excites the RW null space is also not included, keeping $u$ and $\delta \tilde{\omega}_{rw}$ both three-dimensional vectors defined from the set of all input wheel speeds and torques as $u=H_i\tau$ and $\delta \tilde{\omega}_{rw}=H_i\delta \omega_{rw}$, with $H_i\in \mathrm{R}^{3\times N_{rw}}$ the inertial subspace/rigid body rotational control allocation matrix and $N_{rw}$ the number of RWs.
Under the additional assumptions of slow slewing speeds and near constant RW speed, the dynamic constraints are linearized to obtain a convex quadratic program formulation of the guidance algorithm. 
To impose the RW velocity constraints, their velocity is reconstructed from the inertial subspace of the RWs, assuming the null space is always zero.
The cost function is defined solely as a quadratic function of the difference in attitude with respect to the target and the angular velocity. The weights of this function are tuned to avoid underdamped or overdamped slewing profiles.
The refined output of this algorithm for slew maneuvers is a simple, dynamically consistent bang-coast-bang profile that is commonly adopted in ACS for large angle maneuvers \cite{ponche2023guidance}.
Some margin is kept between the real torque and global RW speed constraints and the constraints imposed on the guidance algorithm, as the controller is not designed to reason around constraints.

\subsection{Linear Quadratic Gaussian Controller} 

 The selected controller is the Linear Quadratic Gaussian (LQG), which is the combination of the Linear Quadratic Regulator (LQR) state controller and the Kalman Filter state estimator. Both are the solution to an optimization problem that finds the optimal controller that minimizes a quadratic cost function with respect to the state, controls and the process and measurement noise of the estimator. The LQR controller gain is obtained as the solution to a dynamic programming backward Ricatti recursion which, for a strictly convex quadratic cost function, and the linear dynamic system matrices $(A,B)$ being stabilizable and $(A,Q)$ detectable, converges to a constant value towards infinity. Similarly, the Kalman Filter converges to a steady state gain towards infinity if the covariance of the process and measurement noise $W$ and $V$ are positive semidefinite, system matrices $(C,A)$ are detectable and the $(A,W)$ pair is stabilizable.
The LQG then achieves a steady-state behavior with a linear estimator and controller gain, forming a linear controller. \\

The system dynamics described in Eq. \eqref{eq:sysdyn} are not fully stabilizable and detectable. The unstabilizable subspace considers the total angular momentum of the system and the translational rigid body states, with the latter being also undetectable. The controller is designed only for the stabilizable and detectable subspace of the system. The LQG states therefore exclude the rigid body translational states. The RW are decomposed into the inertial and null space, with the former removed from the controller states but not the estimator. The controller is therefore able to control the attitude of the system without considering the total inertial angular momentum of the wheels. The null space of the wheels is used to help suppress the vibrations of the system in the distributed case study.

\newpage 


\section{Simulation Results}\label{sec:simres}










Two common mission scenarios for high accuracy pointing missions are studied: a coarse pointing slew maneuver and steady-state fine attitude pointing. Different sensor models and assumptions are considered in both scenarios.

\subsection{Slew Maneuver}

\begin{figure*}
    \centering
    \begin{subfigure}[b]{0.32\textwidth}
        \centering
        \includegraphics[width=\textwidth,valign=b]{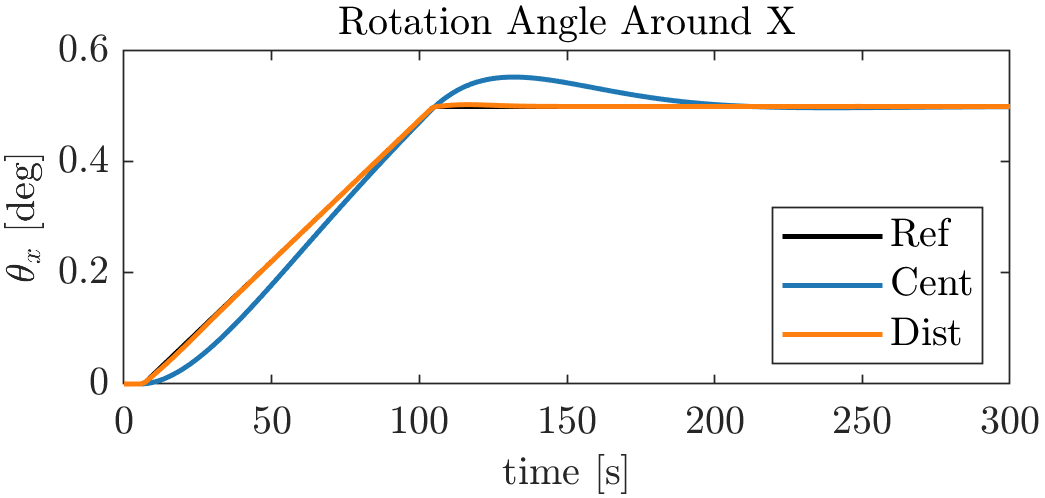}
        \caption[]%
        {{\small Slew Maneuver attitude profile.}} 
        \label{fig:slewatt}
    \end{subfigure}
    \hfill
    \begin{subfigure}[b]{0.32\textwidth}  
        \centering 
        \includegraphics[width=\textwidth]{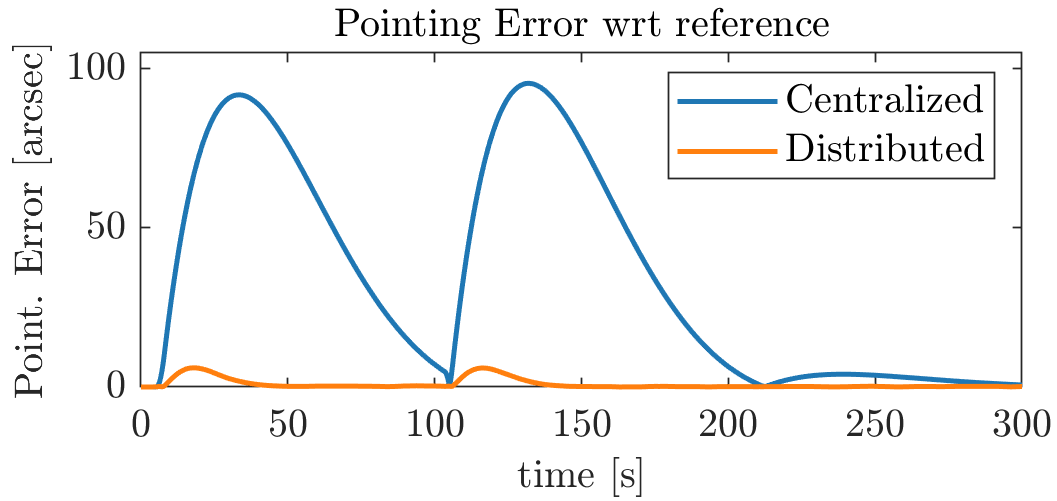}
        \caption[]%
        {{\small Slew Maneuver attitude tracking error.}}    
        \label{fig:slewatterr}
    \end{subfigure}
    \hfill
    \begin{subfigure}[b]{0.32\textwidth}  
        \centering 
        \includegraphics[width=\textwidth]{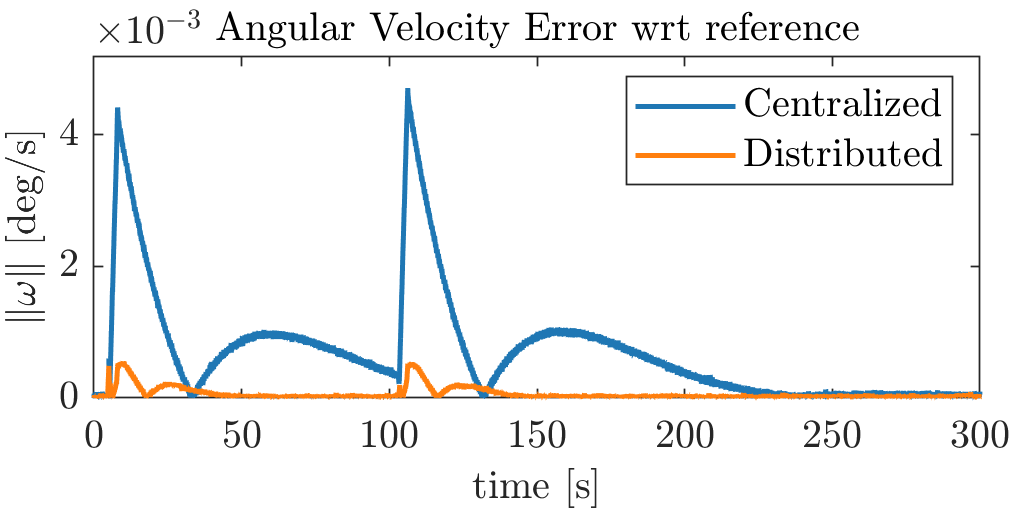}
        \caption[]%
        {{\small Slew Maneuver angular velocity tracking error.}}    
        \label{fig:slewomegaerr}
    \end{subfigure}
    \vskip\baselineskip
    \begin{subfigure}[b]{0.32\textwidth}   
        \centering 
        \includegraphics[width=\textwidth,valign=b]{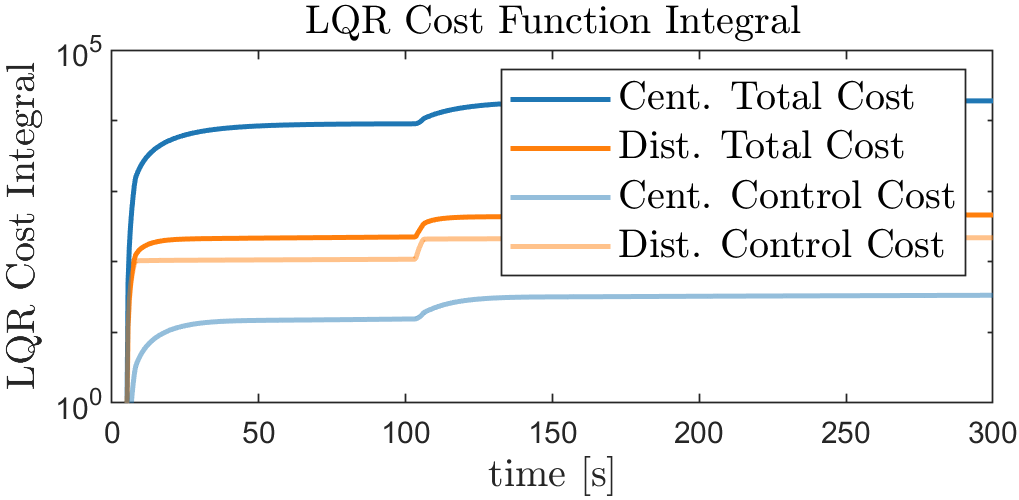}
        \caption[]%
        {{\small Integral of the Controller Cost Function, and the control-only term of the cost function.}}    
        \label{fig:slewlqrcost}
    \end{subfigure}
    \hfill
    \begin{subfigure}[b]{0.32\textwidth}   
        \centering 
        \includegraphics[width=\textwidth,valign=b]{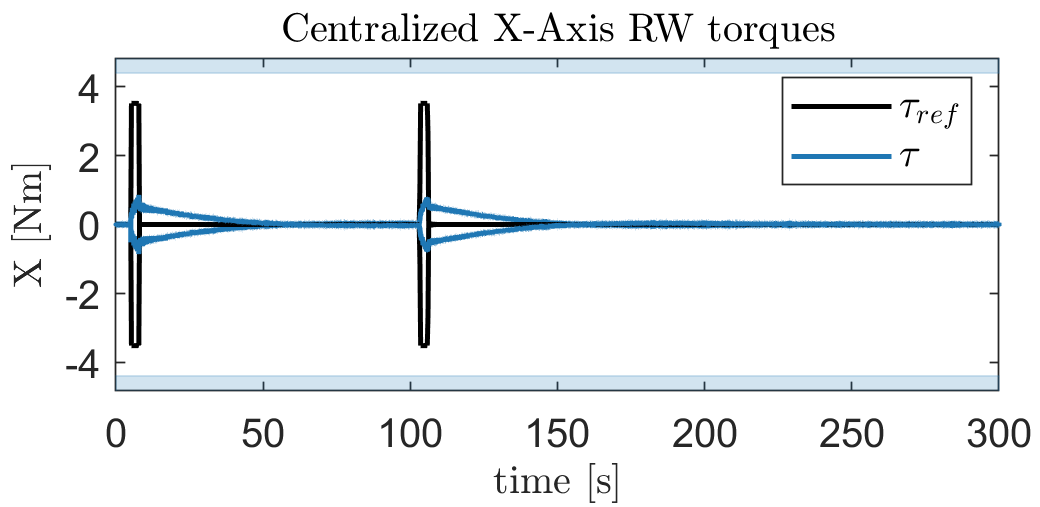}
        \caption[]%
        {{\small Torque of the X-axis aligned RW in the centralized configuration.}}    
        \label{fig:slewcenttau}
    \end{subfigure}
    \hfill
    \begin{subfigure}[b]{0.32\textwidth}  
        \centering 
        \includegraphics[width=\textwidth,valign=b]{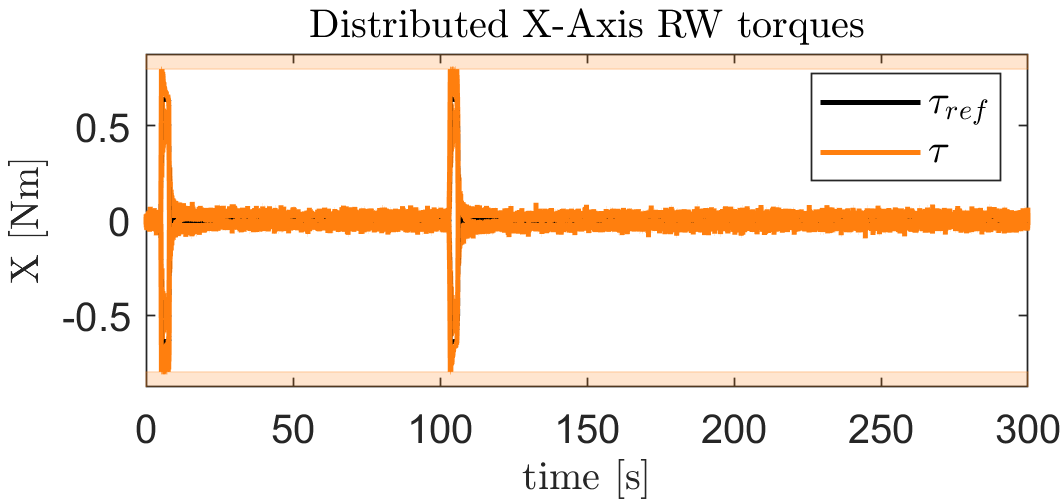}
        \caption[]%
        {{\small Torque of the X-axis aligned RW in the distirbuted configuration.}}    
        \label{fig:slewdisttau}
    \end{subfigure}
    \vskip\baselineskip
    \begin{subfigure}[b]{0.32\textwidth}   
        \centering 
        \includegraphics[width=\textwidth]{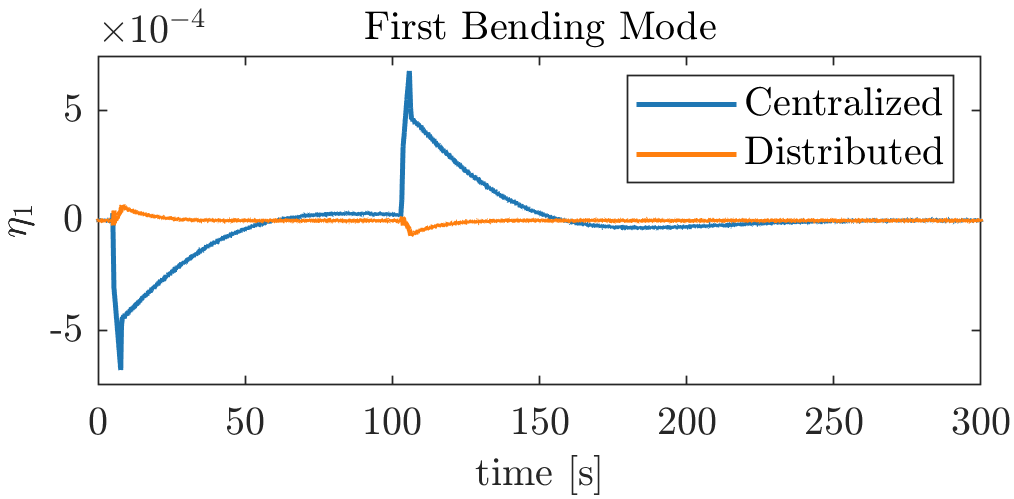}
        \caption[]%
        {{\small First bending mode.}}    
        \label{fig:slewfstbendd}
    \end{subfigure}
    \hfill
    \begin{subfigure}[b]{0.32\textwidth}   
        \centering 
        \includegraphics[width=\textwidth]{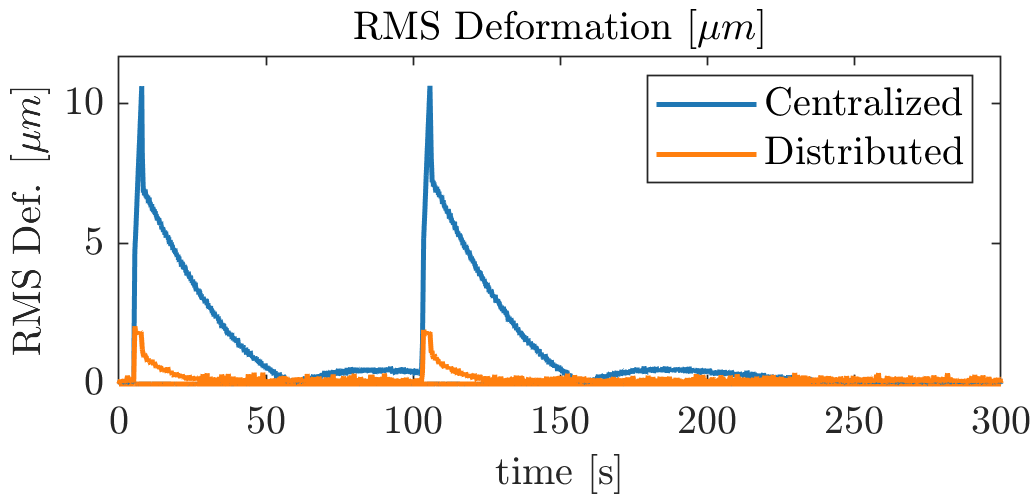}
        \caption[]%
        {{\small RMS Deformation of the structure.}}    
        \label{fig:slewrmsdef}
    \end{subfigure}
    \hfill
    \begin{subfigure}[b]{0.32\textwidth}  
        \centering 
        \includegraphics[width=\textwidth]{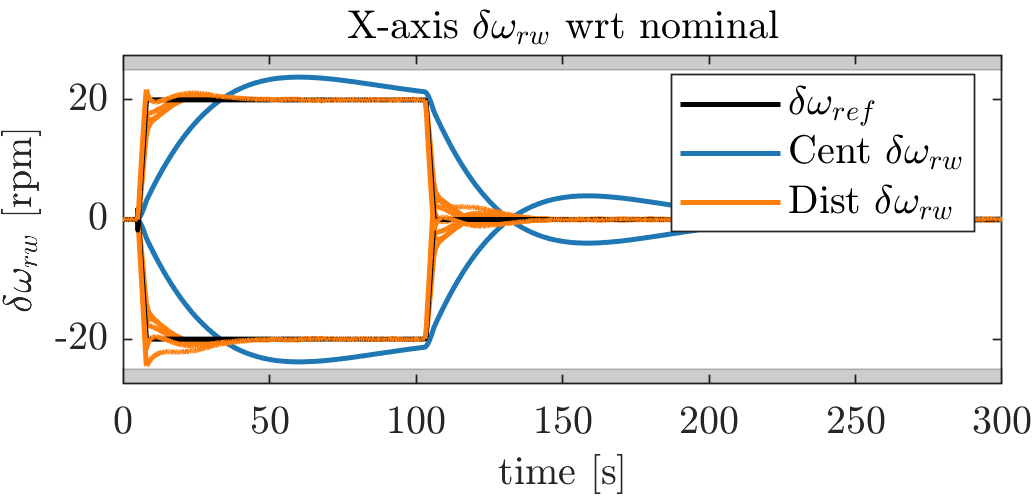}
        \caption[]%
        {{\small RW Speed with respect to their nominal speed.}}    
        \label{fig:slewdwrw}
    \end{subfigure}
    \caption[ The average and standard deviation of critical parameters ]
    {\small Comparison of the distributed and the centralized configuration performances in a slewing maneuver.} 
    \label{fig:results}
\end{figure*}

Both the centralized and distributed configurations are tasked with slewing towards a target attitude that is rotated approximately half a degree around the x-axis. Due to the tight RW speed constraints, the maximum slew speed for this case study mission is $0.25$ degree per minute on each axis, leading to a $\sim 2$ minute slew. A coarse pointing mode is used, with less accurate sensors and in which the RW disturbances are not considered.
A single run of 300 seconds is considered. The variance of the sensor noise is known and used to determine the estimator gain. The simulation parameters are listed in table \ref{tab:simparams} of appendix \ref{app:sysparam}. The control cost Hessian $R$ of the controller objective function is set lower in the centralized case than in the distributed case by the ratio of the number of actuators to match the cost of applying the same torque a single large actuator and multiple smaller ones ($\|R^{1/2}N_u u\|_2^2>N_u \|R^{1/2}u\|_2^2,\forall N_u>1$). 
The guidance algorithm is allowed to produce a profile to within $80\%$ of the torque and wheel speed limits. The results are shown in Fig. \ref{fig:results} and Table \ref{tab:res}.


The controller does not surpass the RW speed constraints during the maneuver, but the controller does reach the torque limit in the distributed case when initiating the slew.

Figure \ref{fig:slewatt} and \ref{fig:slewatterr} shows the distributed configuration tracking more accurately the slew profile than the centralized configuration. The settling time to steady state error of the distributed configuration is around $\sim 30\SI{}{\second}$, while the centralized setting takes over $\sim 100 \SI{}{\second}$ to converge. 

Figure \ref{fig:slewdwrw} shows the X-axis RW angular velocity relative to the respective nominal velocity and the guidance reference. The distributed scenario not only converges more quickly to the reference, but is also able to use the null space of the wheels to stabilize the system. The first bending mode $\eta_1$ is shown in Fig. \ref{fig:slewfstbendd}, the shape of which is shown in Fig. \ref{fig:scmodes}. This mode is tightly coupled with the rotation around the X-axis and shows greater strain when the actuation is limited to the bus. To evaluate the deformation along the entire structure, the quantity $\|\eta\|_2/\sqrt{m_{sc}}$ is used in Fig. \ref{fig:slewrmsdef}, representing the root-mean squared (RMS) deformation/displacement of the structure averaged over its surface. Overall, the structure deforms less during the maneuver when the actuators are distributed throughout the structure. The LQR cost function integral, defined as the sum of quadratic costs on the states and control inputs, is represented in Fig. \ref{fig:slewlqrcost}. While the distributed configuration grows a larger control cost, its total cost is substantially lower when the state error cost is considered. 


\subsection{Fine-pointing}
\begin{figure*}
    \centering
    \begin{subfigure}[b]{0.48\textwidth}
        \centering
        \includegraphics[width=\textwidth]{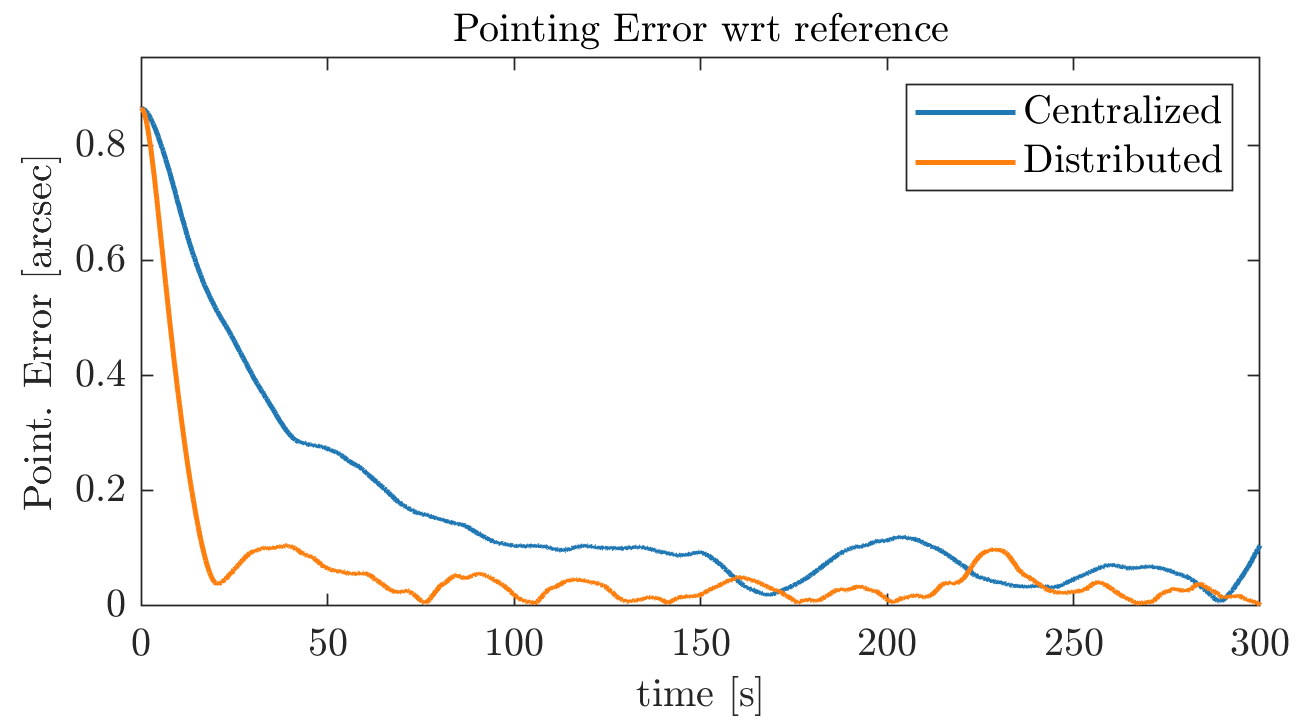}
        \caption[]%
        {{\small Attitude Pointing Error.}}    
        \label{fig:resg}
    \end{subfigure}
    \hfill
    \begin{subfigure}[b]{0.48\textwidth}  
        \centering 
        \includegraphics[width=\textwidth]{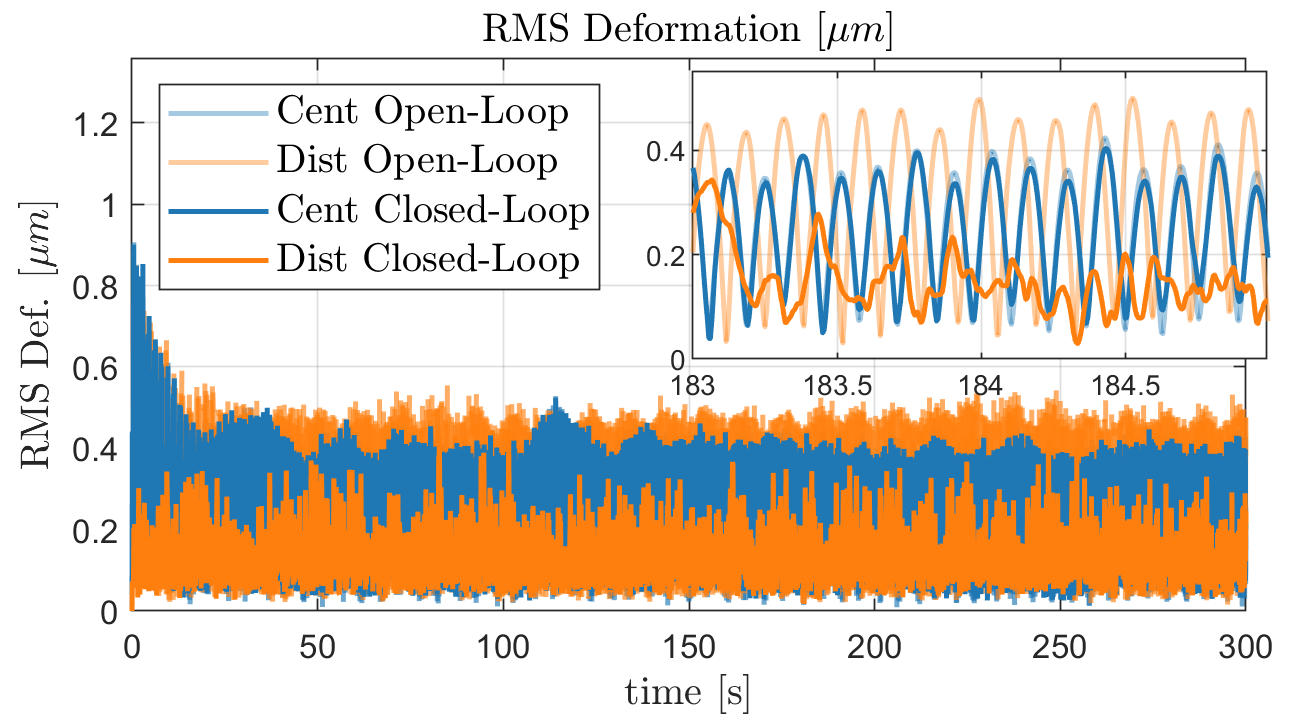}
        \caption[]%
        {{\small Zoom-in of the RMS deformation. Open-loop simulations without vibration suppression show the magnitude of excitation of the structure from the disturbances of the wheels in both configurations.}}    
        \label{fig:resh}
    \end{subfigure}
    \caption[ The average and standard deviation of critical parameters ]
    {\small Comparison of the distributed and the centralized actuator configurations in a fine pointing scenario.} 
    \label{fig:fpresults}
\end{figure*}

In this scenario, a comparison between the performance of the centralized and distributed configurations in a steady-state pointing scenario is also explored. The attitude vision-based sensor is assumed to be a more accurate Fine Guidance Sensor (FGS) instead of the SST used in the slew maneuver. The actuator disturbance models are considered.
The simulation is initialized in a random orientation with a normal distribution of standard deviation $0.1 \SI{}{\arcsecond}$ away from the target, close enough that a guidance profile is not necessary. The ability of the controller to actively reject the disturbances of the RW and achieve fine pointing is compared in both scenarios in a single $300\SI{}{\second}$ run. The results are shown in Fig. \ref{fig:fpresults} and Table \ref{tab:res}.

In Fig. \ref{fig:resg}, the pointing error over time is shown for both configurations. As in the slewing maneuver, the centralized takes longer to converge, after which the pointing error converges to a steady state similar with the distributed configuration. Fig.\ref{fig:resh} shows a short two second span of the RMS deformation of the structure for both configurations, for both open and closed-loop simulations. The open-loop simulations see the SC being initialized in the target attitude, with the dynamics being solely excited by the RW disturbances. These results are shown to illustrate that distributing the momentum actuators makes the structure more sensitive to the disturbances generated by them than when these are centralized in the bus. With the high-bandwidth controller, the distributed scenario manages to significantly attenuate the vibrations induced by the actuators. In the centralized scenario, the controller does not significantly suppress the vibrations induced by the RWs, only having controllability over the modes excited by the torque/dynamic imbalance perturbations but not the force/static imbalance perturbations. The controller of the distributed scenario has the disadvantage of using more torque than the centralized scenario, approximately twice the total torque between all actuators over a given period. This is consistent with what is shown in the slewing scenario in Fig. \ref{fig:slewlqrcost}.
\begin{table}[H]
\captionsetup{justification=centering}
\caption{Performance metrics of the centralized and distributed configurations (CL = closed-loop, OL = open-loop).}\label{tab:res}
\resizebox{\columnwidth}{!}{%
\begin{tabular}{r|c|cl|cl|c}
\multicolumn{1}{l|}{\multirow{2}{*}{\textbf{}}}                            & \multicolumn{1}{l|}{\multirow{2}{*}{\textbf{Metric}}} & \multicolumn{2}{c|}{\textbf{Cent.}}      & \multicolumn{2}{c|}{\textbf{Dist.}}  & \multirow{2}{*}{\textbf{Units}}    \\
\multicolumn{1}{l|}{}                                                              & \multicolumn{1}{l|}{}                                 & \multicolumn{1}{c|}{\textbf{CL}} & \textbf{OL} & \multicolumn{1}{c|}{\textbf{CL}} & \textbf{OL} &                                 \\ \hline
\multirow{7}{*}{\textbf{\begin{tabular}[c]{@{}r@{}}Slew\\  Man.\end{tabular}}} & \textbf{Total Cost}                                   & \multicolumn{1}{c|}{$1.9\mathrm{e}{4}$}            &             & \multicolumn{1}{c|}{$461$}            &          &                                  \\
    & \textbf{Control Cost}                                   & \multicolumn{1}{c|}{$34$}            &             & \multicolumn{1}{c|}{$219$}            &          &                                   \\
    & \textbf{Set. Time}                                   & \multicolumn{1}{c|}{$\approx 100$}            &             & \multicolumn{1}{c|}{$\approx 30$}            &          &                               \SI{}{\second}     \\
  & \textbf{RMS $\|\delta\theta\|_2$}                                  & \multicolumn{1}{c|}{$49$}            &             & \multicolumn{1}{c|}{$1.8$}            &            &             \SI{}{\arcsecond}      \\
   & \textbf{RMS $\|\delta\omega\|_2$}                              & \multicolumn{1}{c|}{$1.1\mathrm{e}{-3}$}            &             & \multicolumn{1}{c|}{$1.1\mathrm{e}{-4}$}            &         &                    \SI{}{\degree/\second}     \\
    & \textbf{RMS Def.}                            & \multicolumn{1}{c|}{$2.3$}            &             & \multicolumn{1}{c|}{$0.33$}            &          &                                   \SI{}{\micro\metre} \\ \hline
\multirow{4}{*}{\textbf{\begin{tabular}[c]{@{}r@{}}Fine \\ Point.\end{tabular}}} & \textbf{RMS $\|\tau\|_1$}                                   & \multicolumn{1}{c|}{$8.3\mathrm{e}{-2}$}            &             & \multicolumn{1}{c|}{$0.24$}            &       &     \SI{}{\newton\meter}    \\
    & \textbf{RMS $\|\delta\theta\|_2$}                                 & \multicolumn{1}{c|}{$6.8\mathrm{e}{-2}$}            &             & \multicolumn{1}{c|}{$3.8\mathrm{e}{-2}$}            &  &       \SI{}{\arcsecond}      \\        
        & \textbf{RMS $\|\delta\omega\|_2$}                                 & \multicolumn{1}{c|}{$3.0\mathrm{e}{-5}$}            &             & \multicolumn{1}{c|}{$2.5\mathrm{e}{-5}$}            &   &           \SI{}{\degree/\second} \\     
            & \textbf{RMS Def.}                                 & \multicolumn{1}{c|}{$0.27$}            &      $0.27$       & \multicolumn{1}{c|}{$0.15$}            &  $0.33$ &         \SI{}{\micro\metre}   \\     
\end{tabular}
}%
\end{table} 

\section{Conclusions}\label{sec:conc}

A distributed momentum actuator approach was proposed for the control of a large flexible space structure.
The performance of this approach was compared with a centralized RW distribution with a single set of platform momentum actuators. A scenario of coarse pointing stabilization and tracking of a slew maneuver and another of fine pointing were studied. The reaction wheel speeds were boxed into a tight bandwidth with minimal interference with the structural dynamics to minimize the propagation of jitter perturbations from the momentum actuators.
The distributed actuator configuration achieved lower displacement within the structure and faster settling times than the centralized actuator configuration for a similar controller cost function, at the expense of more control action. The flexible structure is more sensitive to the disturbances of the actuators when these are dispersed along the structure, but the added controllability over the bending modes of the structure also allows the control system to suppress their disturbances to a level lower than the one present in the centralized scenario.

\appendices{}              

\section{System Parameters}        
\label{app:sysparam}
The parameters of the models described in this paper are listed in this appendix. 
Table \ref{tab:simparams} lists the parameters of the flexible spacecraft model.

\begin{table}[]
\captionsetup{justification=centering}
\caption{Simulation and System parameters.}\label{tab:simparams}
\centering
\begin{tabular}{c|c|c|c}
\textbf{}                                                             & \textbf{Parameters} & \textbf{Value} & \textbf{Unit} \\ \hline
\textbf{\begin{tabular}[c]{@{}r@{}}Flexible\\ Structure\end{tabular}} & $J_{xx}$,$J_{yy}$    &  $2.2\mathrm{e}{5}$  & \SI{}{\kg\meter^2} \\
\textbf{}                                                             & $J_{zz}$  & $4.4\mathrm{e}{5}$  & \SI{}{\kg\meter^2} \\
\textbf{}                                                             & $J_{xy},J_{xz},J_{yz}$ &    $0$    & \SI{}{\kg\meter^2} \\
\textbf{}                                                             & $m$         &       $4200$      & \SI{}{\kg}    \\
\textbf{}                                                             & $\beta_M$        & $3\mathrm{e}{-2}$ &               \\
\textbf{}                                                             & $\beta_K$        & $2\mathrm{e}{-4}$ &               \\ \hline
\textbf{Large RW}                                                     & $J_{rw}$        &  $4.4$ & \SI{}{\kg\meter^2}  \\
\textbf{}                                                             & $\tau_{max}$ & $4.4$ & \SI{}{\newton\meter} \\ \textbf{}                                                             & $C_1^f$ & $1.2\mathrm{e}{-8}$ & \SI{}{\newton \second^2/\degree^2} \\
\textbf{}  & $C_1^{\tau}$ &  $5\mathrm{e}{-9}$ & \SI{}{\newton\meter\second^2/\degree^2} \\
\textbf{}                                                             & Noise & $1.7\mathrm{e}{-2}$ & \SI{}{\degree/\second\sqrt{\Hz}} \\
\hline 
\textbf{Small RW}                                                     & $J_{rw}$            &  $0.8$  & \SI{}{\kg\meter^2} \\
\textbf{}                                                             &$\tau_{max}$ & $0.8$ & \SI{}{\newton\meter} \\
\textbf{}                                                             & $C_1^f$ & $2.2\mathrm{e}{-9}$ & \SI{}{\newton\second^2/\degree^2} \\
\textbf{}                                                             & $C_1^{\tau}$ & $9\mathrm{e}{-10}$ & \SI{}{\newton\meter\second^2/\degree^2} \\
\textbf{}                                                             & Noise & $1.7\mathrm{e}{-2}$ & \SI{}{\degree/\second\sqrt{\Hz}} \\
\hline
\textbf{SST (Coarse)}                                                      & Noise         &   $10$        & \SI{}{\arcsecond/\sqrt{\Hz}}              \\
\textbf{FGS (Fine)}                                                             & Noise           &   $0.1$      & \SI{}{\arcsecond/\sqrt{\Hz}}              \\ \hline
\textbf{AMU}                                                          & Noise  &  $2\mathrm{e}{-5}$ & \SI{}{\m/\second^{2}\sqrt{\Hz}}  \\ \hline
\textbf{Gyro}                                                         & Noise & $1\mathrm{e}{-5}$ & \SI{}{\degree/\second\sqrt{\Hz}}
\end{tabular}%
\end{table}

\acknowledgments
This work was partially supported by Fundação para a Ciência e a Tecnologia (Portuguese Foundation for Science and Technology) through the Carnegie Mellon Portugal Program under fellowship PRT/BD/154921/2022. It was also supported by LARSyS FCT funding (DOI: 10.54499/LA/P/0083/2020, 10.54499/UIDP/50009/2020, and 10.54499/UIDB/50009/2020). Portions of this research were supported by funding from generous anonymous philanthropic donations to the Steward Observatory of the College of Science at the University of Arizona.

\bibliographystyle{IEEEtran}
\bibliography{IEEEfull}

\thebiography

\begin{biographywithpic}
{Pedro Cachim}{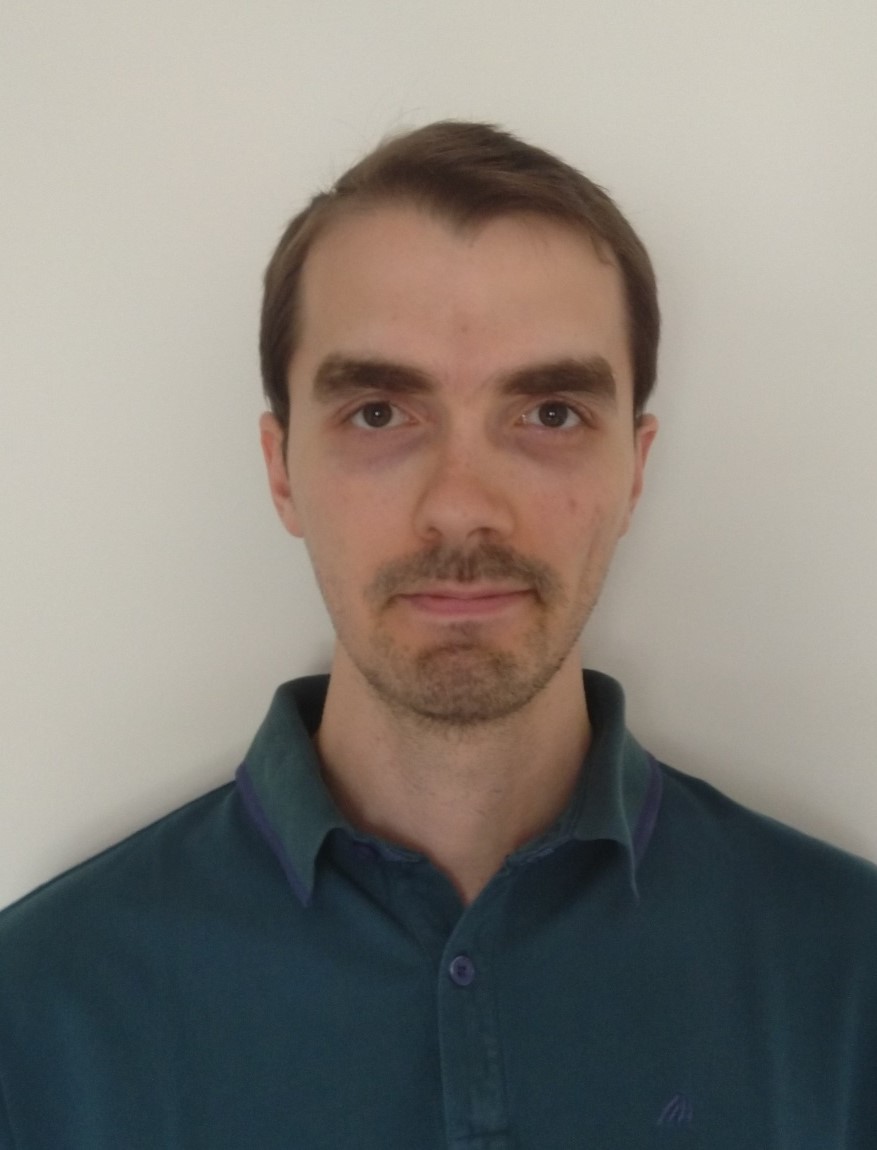}
 received a dual M.S. in Aerospace Engineering from Instituto Superior T\'{e}cnico and ISAE-SUPAERO in 2020 and worked as a GNC engineer at GMV. He is pursuing a dual degree PhD in Electrical and Computer Engineering between the University of Lisbon and Carnegie Mellon University. His research interests include navigation and control for aerospace applications.
\end{biographywithpic}

\begin{biographywithpic}
{Will Kraus}{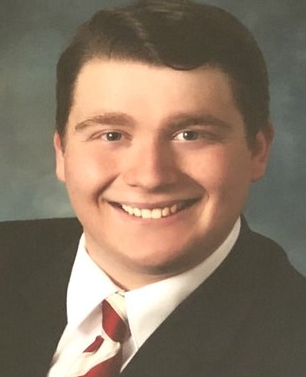}
received his B.S. degree from Pennsylvania State University in Mechanical Engineering in 2023. He is pursuing a M.S. in Mechanical Engineering at Carnegie Mellon University. His research interests include optimal control on real-time hardware systems and data-driven control methods.
\end{biographywithpic} 

\begin{biographywithpic}
{Zach Manchester}{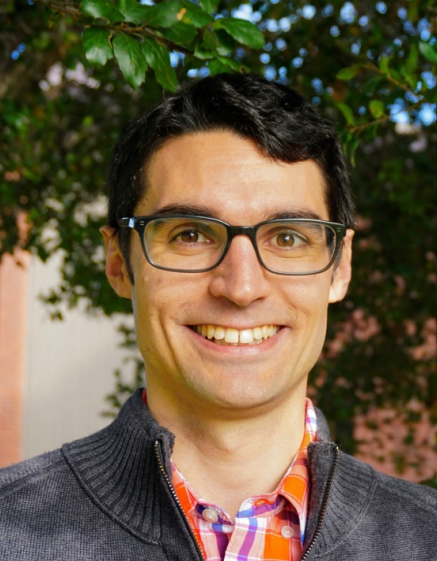}
is an assistant
professor in the Robotics Institute at
Carnegie Mellon University and founder
of the Robotic Exploration Lab. He received a PhD in aerospace engineering
in 2015 and a BS in applied physics
in 2009, both from Cornell University.
His research interests include control
and optimization with applications to
aerospace and robotic systems.
\end{biographywithpic}

\begin{biographywithpic}
{Pedro Louren\c co}{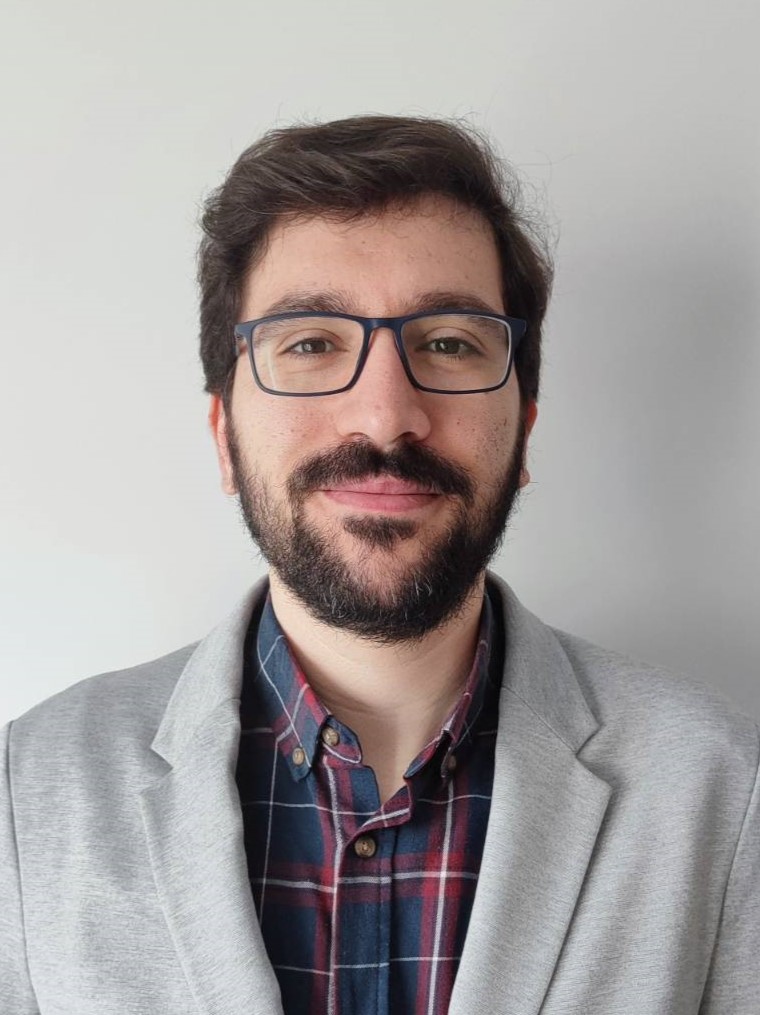}
(PhD) is the head of the Advanced Guidance and Control section within the Flight Segment and Robotics unit of GMV, working on space rendezvous, reusable transportation systems, OSAM, AOCS, etc. As section head, he has led R\&D\&I efforts towards the adoption of robust and optimization-based G\&C and its verification and validation.
\end{biographywithpic} 

\begin{biographywithpic}
{Rodrigo Ventura}{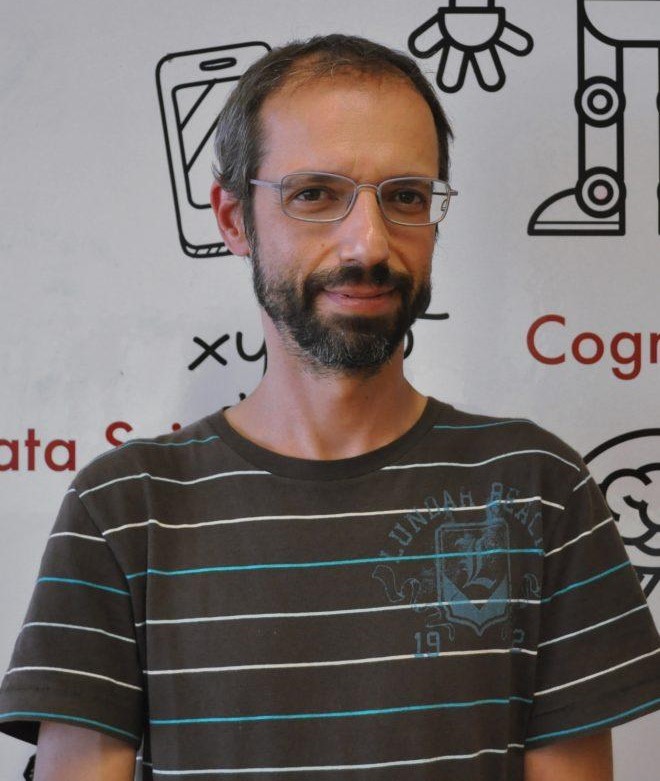}
 (PhD) is a tenured Associate Professor of the Electrical and Computer Engineering Department of Instituto Superior Técnico (IST), University of Lisbon, and a senior researcher of the Institute for Systems and Robotics (ISR-Lisbon). Broadly, his research is focused on the intersection between Robotics and Artificial Intelligence, with particular interest in aerospace applications.
\end{biographywithpic}

\end{document}